\documentclass[seceq,preprint]{ptptex}
\usepackage{wrapft}
\usepackage{graphicx}

\newcommand{\bra}[1]{\langle {#1} |}     
\newcommand{\ket}[1]{| {#1} \rangle}     
\newcommand{\bbra}[1]{\langle\!\langle {#1} |}     
\newcommand{\kket}[1]{| {#1} \rangle\!\rangle}     
\newcommand{\rket}[1]{| {#1} )}     
\newcommand{\maru}[1]{\stackrel{\tiny\circ} {#1}} 
\newcommand{\wtilde}[1]{\widetilde{#1}} 

\newcommand{\kbar}{k \kern -0.5em\raise 0.6ex \hbox{--}}

\def\<{\langle}
\def\>{\rangle}
\def\bsub{\begin{subequations}}
\def\esub{\end{subequations}}
\def\beqn{\begin{eqnarray}}
\def\eeqn{\end{eqnarray}}
\def\b{\begin{equation}}

\title{
A New Boson Realization of Two-Level Pairing Model in Many-Fermion 
System and its Classical Counterpart
}
\subtitle{
Role of Mixed-Mode Coherent State in the Schwinger Boson Representation 
for the $su(2)\otimes su(2)$-Algebra
}    

\author{
Constan\c{c}a {\sc Provid\^encia},$^{1}$
Jo\~ao da {\sc Provid\^encia},$^{1}$\\
Yasuhiko {\sc Tsue}$^{2}$ 
and Masatoshi {\sc Yamamura}$^{3}$
}

\inst{
$^{1}$Departamento de Fisica, Universidade de Coimbra, 3004-516 Coimbra, 
Portugal\\
$^2$Physics Division, Faculty of Science, Kochi University, Kochi 780-8520, 
Japan\\
$^{3}$Faculty of Engineering, Kansai University, Suita 564-8680, Japan
}

\recdate{
\today
}

\abst{
A new method for describing the two-level pairing model in a many-fermion 
system is presented. 
Basic idea comes from the Schwinger boson representation for the 
$su(2)\otimes su(2)$-algebra, in which four kinds of boson operators 
govern the dynamics. 
It is well known that in the original fermion space, it is almost impossible 
to introduce any trial state for the variation except the BCS state in the 
framework of the mean field approximation. 
However, in the Schwinger boson representation, various trial states can be 
defined on the condition that the mean field approximation is useful. 
In these states, the mixed-mode coherent state is mainly discussed 
in this paper. 
}
\begin{document}
\maketitle

\section{Introduction}

The idea of describing the dynamics of many-fermion systems in terms of bosons 
has a long history and is quite natural, since boson degrees of freedom 
correspond obviously to canonical variables of classical mechanics. 
Bosons were already introduced in the collective model of 
Bohr and Mottelson through the quantization of the oscillations 
of a liquid drop to describe excitations of nuclei.\cite{1} 
Sawada derived consistently the so-called random phase approximation 
in a boson 
framework having in mind the electron-gas problem.\cite{2} 
Following his idea, the quasi-particle random phase approximation 
was formulated to describe collective oscillations in spherical nuclei.\cite{3}
With the aid of this method, we can understand the effect of the pairing 
correlations on the collective oscillations. 
Afterward, as a natural stream, the so-called boson expansion methods were 
proposed in various cases.\cite{4,5,6} 
Including these works, many investigations, which are mainly related to 
the Lie algebras, were reviewed in Ref. \citen{7}. 
With the aid of these methods, we can estimate deviations from 
the harmonic oscillations observed in various many-fermion systems. 
The contribution of Marumori and co-workers including one of the 
present authors (M. Y.), who established a boson
mapping 
of operators acting on a fermion Hilbert space into operators on a boson 
Hilbert space,\cite{5} is particularly relevant. 
Of course, this approach was originally conceived to map a fermion pair 
operator into a single boson operator. 
However, the method is very powerful and may be usefully applied 
under more general conditions. 
This point was demonstrated by the present authors with Kuriyama.\cite{8}

Correlations play an important role in quantum many-body systems. 
Their study is a major challenge in various fields of physics. 
The existence of correlations and the emergence of phase transitions are 
interdependent. 
The study of this interdependency in simple models, which has been considered 
by many authors, is very instructive.\cite{9,10} 
In particular, the transition to superconducting or superfluid phase in 
interacting Fermi systems has been attracting the attention of physicists 
since Kamerlingh Onnes made his spectacular discovery in 1911. 
Important aspects of the underlying physics of this phenomena are well 
described by the two-level pairing model, which has been extensively 
investigated by various researchers, for instance, 
in Refs.\citen{11} $\sim$ \citen{13}, we can find 
these works.

It is well known that the building blocks in the two-level pairing model 
are two kinds of Cooper-pair operators and, including the fermion number 
operators, they compose the $su(2)\otimes su(2)$-algebra. 
It is also well known that the BCS theory is a powerful approach for 
this model and we can describe the model in terms of the mean field 
approximation, i.e., the quasi-particle in the Bogoliubov transformation. 
In the framework of this theory, we can find the phase transition point 
to the superconducting phase. 
However, in many-body systems with finite degrees of freedom such as 
nuclei, we cannot observe sharp phase transition. 
For this problem, the BCS theory may be not so powerful as expected, but, 
in the framewrok of the mean field approximation in the fermion space, 
we cannot find an alternative approach. 
Therefore, if we intend to describe this problem, inevitably, 
we must go beyond the mean field approximation.

The promising posibility of performing the bosonization of fermion 
operators by means of a Schwinger-type mapping has also been persued 
by some autheor.\cite{14,15}
Here, following ideas introduced in previous papers of the present 
authors with Kuriyama,\cite{16} we focus on the two-level pairing model 
in the framewrok of the Schwinger boson representation, 
based on the use of four kinds of boson operators. 
Observing that the two-level 
pairing model is essentially governed by the 
$su(2)\otimes su(2)$-algebra and the single $su(2)$-algebra can be 
represented by two kinds of boson operators, 
it is quite natural to describe the two-level 
pairing model in terms of four kinds of bosons.

The aim of this paper is to present a new method for describing the 
two-level pairing model in the Schwinger boson representation. 
The basic idea is similar to that of the BCS theory: 
Introduction of the trial state for the variation in the framework of the 
mean field approximation in the Schwinger boson representation. 
A straightforward transcription from the original fermion representation 
leads to 
the Glauber coherent state. 
However, in the Schwinger boson representation, the product of the 
$su(2)$-generators 
can be diversely transformed by the boson commutation relations. 
Further, we can define, in various forms, the state which corresponds to 
the state $\rket{0}$. 
Here, $\rket{0}$ denotes the state which does not contain any Cooper pair. 
From the above two points, we can construct the trial state for the 
variation in various forms. 
In this paper, we focus on one form, which we will call the 
mixed-mode coherent state. 
With the use of this state, we can construct a classical counterpart of the 
two-level pairing model parametrized in terms of classical canonical 
variables. 
The basic idea is borrowed from Ref.\citen{8}. 
On the other hand, we pay an attention to the fact that our present model 
contains three constants of motion. 
On the basis of this fact, we can derive a disguised form of the two-level 
pairing model expressed in terms of one kind of boson operator. 
The basic idea also comes from Ref.\citen{8}. 
In the sense of the Dirac quantization rule, both are equivalent to 
each other.
Of course, we will discuss some matters which are 
closely related to the mixed-mode coherent state.

After recapitulating the two-level pairing model in the fermion space in \S 2, 
the model is presented in the framework of the Schwinger boson representation 
in \S 3. 
Three constants of motion are defined and following them, the orthogonal set 
is presented. 
In \S 4, a disguised form of the two-level pairing model is formulated 
in the framework of one kind of boson operator. 
The idea comes from the method given in Refs.\citen{5} and \citen{8}. 
Section 5 is a highlight in this paper. 
After the mixed-mode 
coherent state is presented, a classical counterpart of the model is 
formulated. 
In \S 6, the correspondence between the original fermion and the Schwinger 
boson representation is discussed. 
Finally, an idea how to apply the present method is sketched with some 
comments.

\section{Two-level pairing model in fermion space}

The model discussed in this paper is two-level pairing model 
in many-fermion system. 
The two levels are specified by $\sigma=+$ (the upper) and 
$\sigma=-$ (the lower), respectively. 
The difference of the single-particle energies between the two levels 
and the strength of the pairing interaction are denoted as 
$\hbar \epsilon \ (>0)$ and $\hbar^2 G\ (>0)$, respectively. 
The Hamiltonian ${\hat {\cal H}}$ is expressed in the form 
\begin{equation}\label{2-1}
{\hat {\cal H}}=\epsilon({\hat {\cal S}}_0(+)-{\hat {\cal S}}_0(-))
-G({\hat {\cal S}}_+(+)+{\hat {\cal S}}_+(-))
({\hat {\cal S}}_-(+)+{\hat {\cal S}}_-(-))  \ .
\end{equation}
Here, ${\hat {\cal S}}_0(\sigma)$, ${\hat {\cal S}}_+(\sigma)$ and 
${\hat {\cal S}}_-(\sigma)$ are defined as 
\begin{subequations}\label{2-2}
\begin{eqnarray}
& &{\hat {\cal S}}_0(\sigma)
=(\hbar/2)({\hat {\cal N}}_\sigma-\Omega_{\sigma}) \ , 
\label{2-2a}\\
& &{\hat {\cal S}}_+(\sigma)=(\hbar/2){\hat {\cal P}}_\sigma^* \ , \qquad
{\hat {\cal S}}_-(\sigma)=(\hbar/2){\hat {\cal P}}_\sigma \ , 
\qquad\qquad\qquad\qquad\qquad\qquad\qquad\ \ 
\label{2-2b}
\eeqn
\end{subequations}
\vspace{-0.9cm}
\begin{subequations}\label{2-3}
\beqn
& &{\hat {\cal N}}_\sigma
=\sum_{m=-j_\sigma}^{j_{\sigma}}{\hat c}_m^*(\sigma)
{\hat c}_m(\sigma) \ , \qquad \Omega_\sigma=(2j_\sigma+1)/2\ , 
\label{2-3a}\\ 
& &{\hat {\cal P}}_\sigma^*=\sum_{m={-j_\sigma}}^{j_{\sigma}}
{\hat c}_m^*(\sigma)(-)^{j_\sigma-m}{\hat c}_{-m}^*(\sigma)\ , \ 
{\hat {\cal P}}_\sigma=\sum_{m={-j_\sigma}}^{j_{\sigma}}(-)^{j_\sigma-m}
{\hat c}_{-m}(\sigma){\hat c}_{m}(\sigma)\ .\qquad
\label{2-3b}
\end{eqnarray}
\end{subequations}
The set (${\hat c}_m(\sigma),{\hat c}_m^*(\sigma))$ denotes 
fermion operator in the single-particle state\break 
$(\sigma, j_\sigma,m\ ; j_\sigma={\rm half\ integer},\ 
m=-j_\sigma, -j_\sigma+1,\cdots ,
j_\sigma-1,j_\sigma)$. 
The fermion number operator and the fermion pairing operator 
(the Cooper pair) in the 
single-particle state $\sigma$ are denoted by ${\hat {\cal N}}_\sigma$ 
and $({\hat {\cal P}}_\sigma,{\hat {\cal P}}_\sigma^*)$, respectively. 
The set $({\hat {\cal S}}_{\pm,0}(\sigma))$ obeys the 
$su(2)$-algebra, and then, essentially, the presented model is governed by 
the $su(2)\otimes su(2)$-algebra. 
The addition of the two sets also obeys the $su(2)$-algebra: 
\begin{equation}\label{2-4}
{\hat {\cal S}}_{\pm,0}={\hat {\cal S}}_{\pm,0}(+)
+{\hat {\cal S}}_{\pm,0}(-) \ .
\end{equation}

The Hamiltonian (\ref{2-1}) has three constants of motion. 
This can be shown from the following relation: 
\begin{equation}\label{2-5}
[\ {\hat {\mib {\cal S}}}(+)^2\ , \ {\hat {\cal H}}\ ]=
[\ {\hat {\mib {\cal S}}}(-)^2\ , \ {\hat {\cal H}}\ ]=
[\ {\hat {\cal S}}_0\ , \ {\hat {\cal H}}\ ]=0 \ .
\end{equation}
Here, ${\hat {\mib {\cal S}}}(\sigma)^2$ denotes the Casimir operator for 
$({\hat {\cal S}}_{\pm,0}(\sigma))$:
\begin{equation}\label{2-6}
{\hat {\mib {\cal S}}}(\sigma)^2=
{\hat {\cal S}}_0(\sigma)^2+(1/2)\left[
{\hat {\cal S}}_-(\sigma){\hat {\cal S}}_+(\sigma)
+{\hat {\cal S}}_+(\sigma){\hat {\cal S}}_-(\sigma)\right]\ . 
\end{equation}
Since ${\hat {\cal H}}$ contains the term 
$({\hat {\cal S}}_0(+)-{\hat {\cal S}}_0(-))$, 
the Casimir operator for $({\hat {\cal S}}_{\pm,0})$ 
does not commute with ${\hat {\cal H}}$. 
If the seniority numbers for the two single-particle states are zero, we have 
\begin{equation}\label{2-7}
{\rm the\ eigenvalue\ of}\ {\hat {\mib {\cal S}}}(\sigma)^2
=(\hbar\Omega_{\sigma}/2)\left(
\hbar\Omega_{\sigma}/2+\hbar\right)\ .
\end{equation}
The vacuum $\rket{0}$, which obeys ${\hat c}_m(\sigma)\rket{0}=0$, 
is the eigenstate of ${\hat {\mib {\cal S}}}(\sigma)^2$ with the eigenvalue 
(\ref{2-7}), and of course, it obeys 
${\hat {\cal S}}_-(\sigma)\rket{0}=0$ and ${\hat {\cal S}}_0(\sigma)\rket{0}
=-(\hbar\Omega_{\sigma}/2)\rket{0}$. 
Then, we denote $\rket{0}$ as 
\begin{equation}\label{2-8}
\rket{0}=\rket{\Omega_+,\Omega_-}\ .
\end{equation}
The eigenstate of ${\hat {\mib {\cal S}}}(\sigma)^2$ and 
${\hat {\cal S}}_0$ with arbitrary normalization 
is expressed in the following form: 
\begin{equation}\label{2-9}
\rket{(\Omega_+\Omega_-\nu);\kappa}=
\left({\hat {\cal S}}_+(+)\right)^{\kappa/2}
\left({\hat {\cal S}}_+(-)\right)^{(\nu-\kappa)/2}\rket{\Omega_+,\Omega_-}\ .
\end{equation}
The state $\rket{(\Omega_+\Omega_-\nu);\kappa}$ satisfies 
\bsub\label{2-10}
\beqn
& &{\hat {\cal S}}_0\rket{(\Omega_+\Omega_-\nu);\kappa}
=(\hbar/2)(\nu-(\Omega_++\Omega_-))\rket{(\Omega_+\Omega_-\nu);\kappa}\ , 
\label{2-10a}\\
& &{\hat {\cal N}}_+\rket{(\Omega_+\Omega_-\nu);\kappa}
=\kappa\rket{(\Omega_+\Omega_-\nu);\kappa}\ . 
\label{2-10b}
\eeqn
\esub
Of course, $\nu$ and $\kappa$ denote total fermion number and fermion 
number in the state $\sigma=+$, respectively, and they are even integers. 
The quantity $\kappa$ obeys the condition  
\beqn\label{2-11}
& &{\rm (i)}\ 0\leq \kappa \leq \nu\ , \qquad{\rm if}\ 
\nu\leq 2{\rm Min}(\Omega_+ , \Omega_-)\ , \nonumber\\
& &{\rm (ii)}\ 0\leq \kappa \leq 2\Omega_+\ , 
\qquad{\rm if}\ 2\Omega_+\leq \nu 
\leq 2\Omega_- \ , \nonumber\\
& &{\rm (iii)}\ 
\nu-2\Omega_-\leq \kappa \leq \nu\ , \qquad{\rm if}\ 2\Omega_-\leq \nu \leq 2\Omega_+ \ , \nonumber\\
& &{\rm (iv)}\ 
\nu-2\Omega_-\leq \kappa \leq 2\Omega_+ , \qquad{\rm if}\ 
2{\rm Max}(\Omega_+ ,  \Omega_-) \leq \nu \leq 2(\Omega_++\Omega_-)\ . 
\eeqn
Here, if $\Omega_{\sigma} \geq \Omega_{-\sigma}$, 
${\rm Min}(\Omega_+, \Omega_-)=\Omega_{-\sigma}$ and 
${\rm Max}(\Omega_+,\Omega_-)=\Omega_{\sigma}$. 
The condition (\ref{2-11}) can be derived from the inequalities 
$0\leq \kappa\leq 2\Omega_+$ and $0\leq \nu-\kappa \leq 2\Omega_-$, by 
which the state (\ref{2-9}) is governed. 
Then, we can diagonalize ${\hat {\cal H}}$ in terms of a superposition of 
the set 
$\{\rket{(\Omega_+\Omega_-\nu);\kappa}\}$ with a fixed value of 
$(\Omega_+\Omega_-\nu)$. 
In the case which obeys the condition (i) or (ii), 
the eigenstate (\ref{2-9}) can be rewritten as 
\begin{eqnarray}\label{2-12a}
& &\rket{(\Omega_+\Omega_-\nu);\kappa}=
\left({\hat {\cal S}}_+(+){\hat {\cal S}}_-(-)\right)^{\kappa/2}\cdot
\left({\hat {\cal S}}_+(-)\right)^{\nu/2}
\rket{(-) \Omega_+,\Omega_-} \ , \nonumber\\
& &\rket{(-) \Omega_+,\Omega_-}=\rket{\Omega_+,\Omega_-} \ . 
\end{eqnarray}
Here, the normalization is arbitrary. 
Clearly, we have 
\begin{equation}\label{2-13a}
\left({\hat {\cal S}}_+(-){\hat {\cal S}}_-(+)\right)
\cdot\left({\hat {\cal S}}_+(-)\right)^{\nu/2}
\rket{(-) \Omega_+,\Omega_-}=0 \ . 
\end{equation}
The relation (\ref{2-13a}) tells us that in the state 
$({\hat {\cal S}}_+(-))^{\nu/2}\rket{(-) \Omega_+,\Omega_-}$ all fermions 
are occupied in 
the lower level and by the successive operation of $({\hat {\cal S}}_+(+)
{\hat {\cal S}}_-(-))$, we can construct the states in which 
arbitrary numbers of fermions are 
occupied in the upper level. 
This means that the operator $({\hat {\cal S}}_+(+)
{\hat {\cal S}}_-(-))$ plays the same role as that of the raising operators 
in the $su(2)$-algebra. 
In the case of the condition (iii) or (iv), the eigenstate (\ref{2-9}) is 
rewritten as 
\begin{eqnarray}\label{2-12b}
& &\rket{(\Omega_+\Omega_-\nu);\kappa}=
\left({\hat {\cal S}}_+(+){\hat {\cal S}}_-(-)
\right)^{\kappa/2-(\nu/2-\Omega_-)}\cdot
\left({\hat {\cal S}}_+(+)\right)^{\nu/2-\Omega_-}
\rket{(+) \Omega_+,\Omega_-} \ , \nonumber\\
& &\rket{(+) \Omega_+,\Omega_-}=\left({\hat {\cal S}}_+(-)
\right)^{\Omega_-}\rket{\Omega_+,\Omega_-} \ . 
\eeqn
Here, the normalization is arbitrary. 
We have the relation 
\begin{equation}\label{2-13b}
\left({\hat {\cal S}}_+(-){\hat {\cal S}}_-(+)\right)\cdot
\left({\hat {\cal S}}_+(+)\right)^{\nu/2-\Omega_-}
\rket{(+)\Omega_+,\Omega_-}=0 \ .
\end{equation}
This is justified from the relation 
${\hat {\cal S}}_+(-)\rket{(+)\Omega_+,\Omega_-}=
({\hat {\cal S}}_+(-))^{\Omega_-+1}\rket{\Omega_+,\Omega_-}=0$, 
because of the Pauli principle. 
The state (\ref{2-12b}) is for $\nu \geq 2\Omega_-$, and then, the lower state 
is fully occupied and the remaining fermions 
$(\nu-2\Omega_-)$ are occupied in the upper level. 
The other interpretation of the state (\ref{2-12b}) is in parallel to 
that of the state (\ref{2-12a}).

The form (\ref{2-9}) suggests us to introduce the following coherent state: 
\begin{equation}\label{2-14}
\rket{c_0}=N_c\exp(\alpha_+{\hat {\cal S}}_+(+))
\exp(\alpha_-{\hat {\cal S}}_+(-))\rket{\Omega_+,\Omega_-}\ .
\end{equation}
Here, $N_c$ denotes the normalization constant and $\alpha_\sigma$ is 
complex parameter. 
The state (\ref{2-14}) is identical with the trial state for the variation 
used in the BCS 
theory. 
The state $\rket{c_0}$ is constructed in terms of 
an appropriate superposition of the states with 
a fixed value of $(\Omega_+,\Omega_-)$ and different value of $\nu$, 
i.e., fermion number non-conservation. 
For the state (\ref{2-14}), we can introduce the BCS-Bogoliubov transformation 
and the mean field approximation is applicable. 
In addition to the state (\ref{2-14}), we can introduce two 
coherent states, which are suggested by the states (\ref{2-12a}) and 
(\ref{2-12b}):
\bsub\label{2-15}
\begin{eqnarray}
& &\rket{c_-}=N_c\exp[
\beta{\hat {\cal S}}_+(+){\hat {\cal S}}_-(-)]\exp(\gamma
{\hat {\cal S}}_+(-))\rket{(-) \Omega_+,\Omega_-} \ , 
\label{2-15a}\\
& &\rket{c_+}=N_c\exp[
\beta{\hat {\cal S}}_+(+){\hat {\cal S}}_-(-)]\exp(\gamma
{\hat {\cal S}}_+(+))\rket{(+) \Omega_+,\Omega_-} \ . 
\label{2-15b}
\end{eqnarray}
\esub
Here, also, $N_c$ and $(\beta,\gamma)$ denote the normalization constant 
and complex parameter, respectively. 
The states (\ref{2-15a}) and (\ref{2-15b}) are also composed 
of the states with a 
fixed value of $(\Omega_+,\Omega_-)$ and different value of $\nu$. 
The states (\ref{2-15a}) and (\ref{2-15b}) 
may be expected to induce results different 
from those of BCS theory, i.e., 
the form (\ref{2-14}). 
However, it may be impossible to introduce a transformation which makes 
the mean field approximation applicable as it stands. 
The main aim of this paper is to give a form which enables us to describe 
the system in the framework of the mean field approximation.

\section{The Schwinger boson representation for the present system}

Instead of investigating the two-level pairing model in many-fermion system 
in the original fermion space, we describe this model in boson space in 
which the Schwinger boson representation is formulated. 
In \S 6, we show the connection between both forms. 
First, we prepare four kinds of boson operators: 
$({\hat a}_\sigma, {\hat a}_\sigma^*)$ and 
$({\hat b}_\sigma, {\hat b}_\sigma^*)$ $(\sigma=\pm)$. 
Then, we can define the $su(2)$-algebra in the form 
\begin{equation}\label{3-1}
{\wtilde S}_+(\sigma)=\hbar{\hat a}_\sigma^*{\hat b}_\sigma\ , \quad
{\wtilde S}_-(\sigma)=\hbar{\hat b}_\sigma^*{\hat a}_\sigma\ , \quad
{\wtilde S}_0(\sigma)=(\hbar/2)({\hat a}_\sigma^*{\hat a}_\sigma
-{\hat b}_\sigma^*{\hat b}_\sigma)\ . \qquad (\sigma=\pm)
\end{equation}
The Casimir operator ${\wtilde {\mib S}}(\sigma)^2$, which corresponds to 
the form (\ref{2-6}), is expressed as 
\bsub\label{3-2}
\begin{equation}\label{3-2a}
{\wtilde {\mib S}}(\sigma)^2={\wtilde S}(\sigma)
({\wtilde S}(\sigma)+\hbar) \ , \qquad
{\wtilde S}(\sigma)
=(\hbar/2)({\hat a}_{\sigma}^*{\hat a}_{\sigma}+{\hat b}_{\sigma}^*
{\hat b}_{\sigma}) \ . 
\end{equation}
Further, we note that the operator ${\wtilde S}_+(+){\wtilde S}_-(-)$, 
which corresponds to the raising operator 
${\hat {\cal S}}_+(+){\hat {\cal S}}_-(-)$, can be re-formed to 
\begin{equation}\label{3-2b}
{\wtilde S}_+(+){\wtilde S}_-(-)
=(\hbar {\hat a}_+^*{\hat b}_+)(\hbar {\hat b}_-^*{\hat a}_-)
=\hbar^2{\hat a}_+^*{\hat b}_-^*{\hat b}_+{\hat a}_- \ . 
\end{equation}
\esub
In the case of ${\hat {\cal S}}_+(+){\hat {\cal S}}_-(-)$, such a re-formation 
may be impossible. 

The Hamiltonian, which corresponds to the form (\ref{2-1}), can be 
expressed in the form 
\bsub\label{3-3}
\begin{equation}\label{3-3a}
{\wtilde H}=\epsilon({\wtilde S}_0(+)-{\wtilde S}_0(-))
-G({\wtilde S}_+(+)+{\wtilde S}_+(-))({\wtilde S}_-(+)+{\wtilde S}_-(-))\ .
\end{equation}
With the use of ${\wtilde S}_{\pm,0}(\sigma)$ defined in the relation 
(\ref{3-1}), the Hamiltonian (\ref{3-3a}) is written down as 
\beqn\label{3-3b}
{\wtilde H}&=&\epsilon\cdot(\hbar/2)
({\hat a}_+^*{\hat a}_+-{\hat b}_+^*{\hat b}_+-{\hat a}_-^*{\hat a}_-
+{\hat b}_-^*{\hat b}_-)
-G\cdot\hbar^2({\hat a}_+^*{\hat a}_++{\hat a}_-^*{\hat a}_-) \nonumber\\
& &-G\cdot\hbar^2({\hat a}_+^*{\hat a}_+ {\hat b}_+^*{\hat b}_+
+{\hat a}_-^*{\hat a}_-{\hat b}_-^*{\hat b}_-) 
-G\cdot\hbar^2({\hat a}_+^*{\hat b}_-^* {\hat b}_+{\hat a}_-
+{\hat a}_-^*{\hat b}_+^*{\hat b}_-{\hat a}_+) \ . \nonumber\\
& &
\eeqn
\esub
We can see that the following three operators, which are mutually commuted 
with each other, are commuted with the Hamiltonian (\ref{3-3}):
\bsub\label{3-4}
\beqn
& &{\wtilde L}=(\hbar/2)({\hat a}_+^*{\hat a}_+ + {\hat b}_+^*{\hat b}_+) \ , 
\label{3-4a}\\
& &{\wtilde M}=(\hbar/2)({\hat a}_+^*{\hat a}_+ + {\hat a}_-^*{\hat a}_-) \ , 
\label{3-4b}\\
& &{\wtilde T}=(\hbar/2)(-{\hat a}_+^*{\hat a}_+ + {\hat b}_-^*{\hat b}_-
+1) \ . 
\label{3-4c}
\eeqn
Further, for $\hbar{\hat a}_+^*{\hat a}_+$, which commutes with 
${\wtilde L}$, ${\wtilde M}$ and ${\wtilde T}$, but, does not commute with 
${\wtilde H}$, we use the notation 
\begin{equation}\label{3-4d}
{\wtilde K}=\hbar{\hat a}_+^*{\hat a}_+ \ . 
\end{equation}
\esub
As is clear from the relation (\ref{3-2a}), ${\wtilde L}$ denotes 
the magnitude of the $su(2)$-spin for $\sigma=+$. 
Of course, it is positive-definite. 
We can define 
the $su(1,1)$-algebra in the form 
\bsub\label{3-5}
\begin{equation}\label{3-5a}
{\wtilde T}_+=\hbar{\hat a}_+^*{\hat b}_-^*\ , \qquad
{\wtilde T}_-=\hbar{\hat b}_-{\hat a}_+ \ , \qquad
{\wtilde T}_0=(\hbar/2)({\hat a}_+^*{\hat a}_++{\hat b}_-^*{\hat b}_-+1) \ .
\end{equation}
The Casimir operator ${\wtilde {\mib T}}^2$ can be expressed as 
\begin{equation}\label{3-5b}
{\wtilde {\mib T}}^2={\wtilde T}_0^2
-(1/2)\left[{\wtilde T}_-{\wtilde T}_+ + {\wtilde T}_+{\wtilde T}_-
\right]={\wtilde T}({\wtilde T}-\hbar) \ .
\end{equation}
\esub
Therefore, ${\wtilde T}$ can be regarded as the magnitude of the 
$su(1,1)$-spin 
given in the relation (\ref{3-5a}). 
It should be noted that ${\wtilde T}$ is not positive-definite. 
At the present stage, we cannot give any meanings to ${\wtilde M}$ and 
${\wtilde K}$ except the ${\hat a}$-boson number operators. 
The Hamiltonian (\ref{3-3b}) can be expressed in the form 
\beqn\label{3-6}
{\wtilde H}&=&-\left[
\epsilon({\wtilde L}+{\wtilde M}-({\wtilde T}-\hbar/2))
+4G{\wtilde T}{\wtilde M}
\right]\nonumber\\
& &+2\left[\epsilon-G({\wtilde L}+{\wtilde M}-({\wtilde T}-\hbar/2))\right]
{\wtilde K}
+2G{\wtilde K}^2\nonumber\\
& &-G\cdot\hbar^2\left(
{\hat a}_+^*{\hat b}_-^* {\hat b}_+{\hat a}_- + 
{\hat a}_-^*{\hat b}_+^* {\hat b}_-{\hat a}_+\right)\ . 
\eeqn

The Hamiltonian (\ref{3-6}) can be diagonalized in the space spanned 
by the orthogonal set $\{\kket{(tml);k}\}$:
\beqn\label{3-7}
\kket{(tml);k}&=&\left(\sqrt{k!(2t\!-\!1\!+\!k)!(2m\!-\!k)!(2l\!-\!k)!}
\right)^{-1}
\nonumber\\
& &\times
({\hat a}_+^*)^k({\hat b}_-^*)^{2t-1+k}({\hat a}_-^*)^{2m-k}
({\hat b}_+^*)^{2l-k}\kket{0}\ .
\eeqn
Here, $\hbar k$, $\hbar t$, $\hbar m$ and $\hbar l$ denote the 
eigenvalues of the operators ${\wtilde K}$, ${\wtilde T}$, 
${\wtilde M}$ and ${\wtilde L}$, respectively. 

For the quantum numbers, $k$, $t$, $m$ and $l$, we note 
$k=0,1,2,\cdots$, $m=0,1/2,1,\cdots$ and $l=0,1/2,1,\cdots$, but 
$t=1/2,1,3/2, \cdots$ or $0,-1/2, -1,\cdots$. 
Further, $2t-1+k=0,1,2,\cdots$, $2m-k=0,1,2,\cdots$ and 
$2l-k=0,1,2,\cdots$. 
From the above conditions, we obtain 
\begin{equation}\label{3-8}
l=0,\ 1/2,\ 1, \cdots,\qquad m=0,\ 1/2,\ 1,\cdots \ .
\end{equation}
However, concerning $t$ and $k$, we obtain 
\bsub\label{3-9}
\beqn
& &{\rm (i)}\ \ t=1/2,\ 1,\ 3/2, \cdots \ , \qquad
k=0,\ 1,\ 2, \cdots ,\  k^0 \ , 
\label{3-9a}\\
& &{\rm (ii)}\ t=0,\ -1/2,\ -1, \cdots \ , \ -(k^0-1)/2,\qquad
k=1-2t,\ 2-2t,\cdots ,\  k^0 \ , \qquad
\label{3-9b}
\eeqn
\esub
Here, $k^0$ denotes $k^0=2{\rm Min}(l,m)$. 
The diagonalization is performed in the space with a fixed value of $(tml)$, 
and then, one kind of degree of freedom contributes to the diagonalization. 
In the next section, we search a method for this procedure in the case 
(i) shown in the relation (\ref{3-9a}). 
At several places, we will contact with the case (ii) briefly in 
relation to the case (i).

\section{A disguised representation formulated in terms of the MYT boson mapping method}

In order to describe the present system in the framework of one kind 
of degree of freedom, we adopt the basic idea of the MYT boson mapping method, 
which was presented by Marumori, Yamamura (one of the present authors) 
and Tokunaga.\cite{5} 
First, we prepare a boson space spanned by one kind of boson 
operator $({\hat c},{\hat c}^*)$. 
The orthogonal set is given by 
\begin{equation}\label{4-1}
\ket{k}=\left(\sqrt{k!}\right)^{-1}({\hat c}^{*})^k\ket{0}\ .
\end{equation}
Let the state $\ket{k}$ be in one-to-one correspondence with the state 
$\kket{(tml);k}$ shown in the relation (\ref{3-7}): 
\begin{equation}\label{4-2}
\kket{(tml);k}\sim \ket{k} \ . 
\end{equation}
Since 
we investigate the case (i) shown in the relation (\ref{3-9a}), 
$\ket{k}$ has 
its meaning in the case $k\leq k^0$ and we call the space with $k\leq k^0$ 
as the physical space. 
Following the basic idea of the MYT boson mapping method, 
we define the mapping 
operator ${\hat U}$ from the space $\{\kket{(tml);k}\}$ to the physical space 
$\{\ket{k}\}$ as follows: 
\begin{equation}\label{4-3}
{\hat U}=\sum_{k=0}^{k^0}\ket{k}\bbra{(tml);k}\ .
\end{equation}
We can map the state $\kket{(tml);k}$ to $\ket{k}$ by 
\begin{equation}\label{4-4}
{\hat U}\kket{(tml);k}=\ket{k}\ .
\end{equation}
The operator ${\hat U}$ satisfies 
\bsub\label{4-5}
\beqn
& &{\hat U}^{\dagger}{\hat U}=\sum_{k=0}^{k^0}
\kket{(tml);k}\bbra{(tml);k}=1\ , 
\label{4-5a}\\
& &{\hat U}{\hat U}^{\dagger}=\sum_{k=0}^{k^0}
\ket{k}\bra{k}={\hat P}\ . 
\label{4-5b}
\eeqn
\esub
Here, ${\hat P}$ plays a role of the projection to the physical space. 

Any operator ${\wtilde O}$ working in the space $\{\kket{(tml);k}\}$ 
can be mapped in the form 
\begin{equation}\label{4-6}
{\hat O}={\hat U}{\wtilde O}{\hat U}^{\dagger} \ . 
\end{equation}
For example, we have 
\bsub\label{4-7}
\beqn
& &{\hat U}{\wtilde L}{\hat U}^{\dagger}=L{\hat P} \ , \qquad
{\hat U}{\wtilde M}{\hat U}^{\dagger}=M{\hat P} \ , \qquad
{\hat U}{\wtilde T}{\hat U}^{\dagger}=T{\hat P} \ , 
\label{4-7a}\\
& &\ \ L=\hbar l\ , \qquad M=\hbar m\ , \qquad T=\hbar t\ . 
\label{4-7b}
\eeqn
\esub
Further, ${\wtilde K}$ is mapped as 
\begin{equation}\label{4-8}
{\hat U}{\wtilde K}{\hat U}^{\dagger}=\hbar{\hat c}^*{\hat c}
{\hat P} \ . 
\end{equation}
As was discussed by Marshalek,\cite{7} 
the operator ${\hat P}$ appearing in the 
expressions (\ref{4-7a}) and (\ref{4-8}) may be omitted if 
restricted to the physical space. 
Our main interest is in the case $\hbar^2{\hat a}_+^*{\hat b}_-^* 
{\hat b}_+{\hat a}_-$ appearing in the last term of the Hamiltonian 
(\ref{3-6}).
In this case, we have 
\beqn\label{4-9}
& &{\hat U}{\hat a}_+^*{\hat b}_-^*{\hat b}_+{\hat a}_-{\hat U}^{\dagger}
\nonumber\\
&=&\sum_{k=0}^{k^0}\sqrt{k+1}\sqrt{2t+k}\sqrt{2m-k}\sqrt{2l-k}\cdot
\ket{k+1}\bra{k}\nonumber\\
&=&\sum_{k=0}^{k^0}\sqrt{k+1}\sqrt{2t+k}\sqrt{2m-k}\sqrt{2l-k}\cdot
\frac{1}{\sqrt{k+1}}{\hat c}^*\ket{k}\bra{k}\nonumber\\
&=&{\hat c}^*\cdot\sqrt{2t+{\hat c}^*{\hat c}}
\sqrt{2m-{\hat c}^*{\hat c}}\sqrt{2l-{\hat c}^*{\hat c}}\cdot{\hat P} \ .
\eeqn
Then, we obtain 
\beqn\label{4-10}
& &{\hat U}\hbar^2
{\hat a}_+^*{\hat b}_-^*{\hat b}_+{\hat a}_-{\hat U}^{\dagger}
\nonumber\\
&=&\sqrt{\hbar}{\hat c}^*\cdot\sqrt{2T+\hbar{\hat c}^*{\hat c}}
\sqrt{2M-\hbar{\hat c}^*{\hat c}}\sqrt{2L-\hbar{\hat c}^*{\hat c}}
\cdot{\hat P} \ .
\eeqn
We can see that the form (\ref{4-10}) is a mixture of the 
Holstein-Primakoff representations of the $su(2)$- and the $su(1,1)$-algebra 
under the correspondence 
$\sqrt{\hbar}{\hat a}_+^*\rightarrow \sqrt{\hbar}{\hat c}^*$, 
$\sqrt{\hbar}{\hat b}_-^*\rightarrow \sqrt{2T+\hbar{\hat c}^*{\hat c}}$, 
$\sqrt{\hbar}{\hat a}_-\rightarrow \sqrt{2M-\hbar{\hat c}^*{\hat c}}$ and 
$\sqrt{\hbar}{\hat b}_+\rightarrow \sqrt{2L-\hbar{\hat c}^*{\hat c}}$. 
In this case, also we may omit ${\hat P}$. 
Under the above idea, the Hamiltonian (\ref{3-6}) is mapped to the 
following form:
\beqn
{\hat H}&=&
-\left[\epsilon(L+M-(T-\hbar/2))+4GTM\right]\nonumber\\
& &+2\left[\epsilon-G(L+M-(T-\hbar/2))\right]{\hat K}
+2G{\hat K}^2\nonumber\\
& &-G\biggl[\sqrt{\hbar}{\hat c}^*\cdot \sqrt{2T+{\hat K}}\sqrt{2M-{\hat K}}
\sqrt{2L-{\hat K}}\nonumber\\
& &\qquad\quad 
+\sqrt{2L-{\hat K}}\sqrt{2M-{\hat K}}\sqrt{2T+{\hat K}}\cdot \sqrt{\hbar}
{\hat c}\biggl] \ , 
\label{4-11}\\
{\hat K}&=&\hbar{\hat c}^*{\hat c}\ .
\eeqn
We can see that our system can be described in terms of one 
kind of degree of freedom $({\hat c}, {\hat c}^*)$ under a fixed 
value of $(T,M,L)$. 
The above form is unchanged in the case (ii) shown in the 
relation (\ref{3-9b}). 
It should be noted that in this case $T$ is negative.

Let us replace the operator $({\hat c},{\hat c}^*)$, which is the 
$q$-number, with the $c$-number $(c,c^*)$ regarded as canonical: 
\begin{equation}\label{4-13}
{\hat c}\longrightarrow c\ , \qquad {\hat c}^*\longrightarrow c^*\ . 
\quad ({\hat K}\longrightarrow K)
\end{equation}
Then, we have 
\beqn
{H}&=&
-\left[\epsilon(L+M-(T-\hbar/2))+4GTM\right]\nonumber\\
& &+2\left[\epsilon-G(L+M-(T-\hbar/2))\right]{K}+2GK^2\nonumber\\
& &-G(\sqrt{\hbar}{c}^*+\sqrt{\hbar}c)\sqrt{2T+{K}}\sqrt{2M-{K}}
\sqrt{2L-{K}} \ . 
\label{4-14}
\eeqn
The Hamiltonian (\ref{4-14}) can be rewritten as 
\beqn
{H}&=&
-\left[\epsilon(L+M-(T-\hbar/2))+4GTM\right]\nonumber\\
& &+2\left[\epsilon-G(L+M-(T-\hbar/2))\right]{K}+2GK^2\nonumber\\
& &-2G\sqrt{K(2T+K)(2M-K)(2L-K)}\cos\psi \ . 
\label{4-15}
\eeqn
Here, $\psi$ is defined as 
\begin{equation}\label{4-16}
\sqrt{\hbar}c=\sqrt{K}e^{-i\psi}\ , \qquad 
\sqrt{\hbar}c^*=\sqrt{K}e^{i\psi} \ .
\end{equation}
The quantities $K$ and $\psi$ may be regarded as action and angle variables. 

It may be self-evident that the Hamiltonian $H$ is the classical counterpart 
of the Hamiltonian ${\hat H}$, because we have the following relations 
given below for 
the commutators and the Poisson brackets: 
The pioneering idea was presented by Marshalek and Holzwarth.\cite{17} 
For any functions $f(x)$, $\xi(x)$ and $\eta(x)$ ($g(x)=\xi(x)\eta(x)$) 
for $x$, there exist the relations 
\bsub\label{4-17}
\beqn
& &[\ \sqrt{\hbar}{\hat c}\ , f({\hat K}) \ ]
=\hbar\cdot\sqrt{\hbar}{\hat c}f'({\hat K},\hbar) \ , \nonumber\\
& &[\ \xi({\hat K})\cdot \sqrt{\hbar}{\hat c}\ , \ \sqrt{\hbar}{\hat c}^*
\cdot \eta({\hat K})\ ]=
\hbar\cdot\left[g({\hat K})+{\hat K}g'({\hat K},\hbar)\right] \ , 
\label{4-17a}\\
& &[\ \sqrt{\hbar}{c} \ , \ f({K}) \ ]_P
=(-i)\cdot\sqrt{\hbar}cf'(K) \ , \nonumber\\
& &[\ \xi({K})\cdot \sqrt{\hbar}{c}\ , \ \sqrt{\hbar}{c}^*
\cdot \eta({K})\ ]_P=
(-i)\left[g({K})+{K}g'({K})\right] \ , 
\label{4-17b}
\eeqn
\esub
Here, $f'({\hat K},\hbar)$ and $g'({\hat K},\hbar)$ denote the differences 
of the first order for $f({\hat K})$ and $g({\hat K})$ with respect to 
$\hbar$. 
The difference for $F({\hat K})$ is defined as 
\begin{equation}\label{4-18}
F'({\hat K},\hbar)=(F({\hat K})-F({\hat K}-\hbar))/\hbar \ .
\end{equation}
Of course, $f'(K)$ and $g'(K)$ denote the differentials of the 
first order for $f(K)$ and $g(K)$. 
The Poisson bracket for $A$ and $B$ is defined as 
\begin{equation}\label{4-19}
[\ A\ , \ B\ ]_P=
\frac{1}{i\hbar}\left(\frac{\partial A}{\partial c}
\frac{\partial B}{\partial c^*}-\frac{\partial B}{\partial c}
\frac{\partial A}{\partial c^*}\right)
=\frac{\partial A}{\partial \psi}\frac{\partial B}{\partial K}
-\frac{\partial B}{\partial \psi}\frac{\partial A}{\partial K}\ .
\end{equation}
In the sense of Dirac, the 
expressions (\ref{4-17}) and (\ref{4-18}) show that 
the quantum and the classical system correspond to each other. 
The term, which are of the higher order than $O(\hbar^0)$, appearing 
on the right-hand sides of the relation (\ref{4-17a}) 
automatically disappear in the classical form (\ref{4-17b}). 
They express the quantal fluctuations. 
The term (\ref{4-10}) which is contained in the Hamiltonian (\ref{4-11}) 
is in the above situation.

\section{Mixed-mode coherent state inducing the classical counterpart}

In \S 4, we derived a disguised form of the two-level pairing model 
in the Schwinger boson representation and its classical counterpart. 
The basic ideas were the use of the MYT boson mapping method and 
the replacement of the $q$-number with the corresponding $c$-number. 
In this section, we show that there exists a wave packet, which we call 
a mixed-mode coherent state, inducing the classical counterpart of 
the disguised form. 
The idea is borrowed from that presented by the present authors with 
Kuriyama and it was applied to the cases of the $su(2)$- and 
the $su(1,1)$-algebra.\cite{8}

First, we note the form (\ref{3-7}). 
In the case (\ref{3-9a}), the state (\ref{3-7}) 
is rewritten as 
\bsub\label{5-1}
\beqn
& &\kket{(tml);k}=
\sqrt{\frac{(2t-1)!}{(2m)!(2l)!}}\sqrt{\frac{(2m-k)!(2l-k)!}{k!(2t-1+k)!}}
\cdot({\hat a}_+^*{\hat b}_-^*{\hat b}_+{\hat a}_-)^k\kket{(-)tml} \ , 
\label{5-1a}\\
& &\kket{(-)tml}=
\left(\sqrt{(2t-1)!(2m)!(2l)!}\right)^{-1}
\cdot
({\hat b}_-^*)^{2t-1}({\hat a}_-^*)^{2m}({\hat b}_+^*)^{2l}\kket{0} \ . \quad
\label{5-1b}
\eeqn
\esub
The state $\kket{(-)tml}$ obeys 
\begin{equation}\label{5-2}
({\hat a}_-^*{\hat b}_+^*{\hat b}_-{\hat a}_+)\kket{(-)tml}=0 \ . 
\end{equation}
We can see, in the relations (\ref{5-1}) and (\ref{5-2}), the following 
two points: 
(1) The state $\kket{(-)tml}$ 
can be regarded as the state similar to the states with the minimum weight 
in the $su(2)$- and the $su(1,1)$-algebra and (2) 
${\hat a}_+^*{\hat b}_-^*{\hat b}_+{\hat a}_-$ plays the same role as 
that of the raising operators in the $su(2)$- and the $su(1,1)$-algebra. 
Then, following the idea for constructing wave packets inducing the 
classical counterparts of the $su(2)$- and the $su(1,1)$-algebra 
developed by the present authors, 
we set up a wave packet in the form 
\beqn\label{5-3}
\kket{c_-}&=&N_c\exp\left(\frac{\hbar}{A_-B_+}\frac{V_+}{U_+}
{\hat a}_+^*{\hat b}_-^*{\hat b}_+{\hat a}_-\right)\nonumber\\
& &\times \exp\left(\frac{1}{\sqrt{\hbar}}\frac{W_-}{U_+}{\hat b}_-^*\right)
\exp\left(\frac{A_-}{\sqrt{\hbar}}{\hat a}_-^*\right)
\exp\left(\frac{B_+}{\sqrt{\hbar}}{\hat b}_+^*\right)\kket{0} \ . 
\eeqn
Here, $W_-$, $V_+$, $A_-$ and $B_+$ denote complex parameters and the 
parameter $U_+$ and the normalization constant $N_c$ are given as 
\beqn
& &U_+=\sqrt{1+|V_+|^2} \ , 
\label{5-4}\\
& &N_c=(U_+)^{-1}\exp\left(-\frac{1}{2\hbar}(|W_-|^2+|A_-|^2+|B_+|^2)\right) 
 \ . 
 \label{5-5}
\eeqn
The state $\kket{c_-}$ can be rewritten as 
\beqn\label{5-6}
\kket{c_-}&=&N_c\exp\left(
\frac{V_+}{U_+}{\hat a}_+^*{\hat b}_-^*+\frac{1}{\sqrt{\hbar}}\frac{W_-}{U_+}
{\hat b}_-^*\right)
\cdot \exp\left(\frac{A_-}{\sqrt{\hbar}}{\hat a}_-^*
+\frac{B_+}{\sqrt{\hbar}}{\hat b}_+^*\right)\kket{0}\ . 
\eeqn
It may be interesting to see that the state $\kket{c_-}$ is the product of 
the coherent states of the $su(2)$- and the $su(1,1)$-spin system, and then, 
we call 
it the mixed-mode coherent state. 
For the state $\kket{c_-}$, we can introduce the boson operators 
$({\hat \alpha}_+,{\hat \alpha}_+^*)$, 
$({\hat \beta}_-,{\hat \beta}_-^*)$, 
$({\hat \alpha}_-,{\hat \alpha}_-^*)$ and 
$({\hat \beta}_+,{\hat \beta}_+^*)$ 
satisfying the condition 
\bsub\label{5-7}
\begin{equation}\label{5-7a}
{\hat \alpha}_+\kket{c_-}={\hat \beta}_-\kket{c_-}=
{\hat \alpha}_-\kket{c_-}={\hat \beta}_+\kket{c_-}=0\ . 
\end{equation}
The relation to the original one is given in the following transformation: 
\begin{eqnarray}\label{5-7b}
& &{\hat a}_+=U_+{\hat \alpha}_+ + V_+{\hat \beta}_-^*
+\frac{V_+W_-^*}{\sqrt{\hbar}} \ , \qquad 
{\hat b}_-=V_+{\hat \alpha}_+^* + U_+{\hat \beta}_-
+\frac{U_+W_-^*}{\sqrt{\hbar}} \ , \nonumber\\
& &{\hat a}_-={\hat \alpha}_-+\frac{A_-}{\sqrt{\hbar}} \ , \qquad
{\hat b}_+={\hat \beta}_+ + \frac{B_+}{\sqrt{\hbar}} \ . 
\end{eqnarray}
\esub
The transformation (\ref{5-7b}) permits us to apply the 
mean field approximation to the present form. 
The aim of this section is to show that the expectation value of 
${\wtilde H}$ given in the relation (\ref{3-6}) for $\kket{c_-}$ 
becomes the Hamiltonian in the classical counterpart 
(\ref{4-14}) or (\ref{4-15}). 

First, we calculate the expectation values of 
$\hbar{\hat a}_+^*{\hat a}_+$, 
$\hbar{\hat b}_-^*{\hat b}_-$, $\hbar{\hat a}_-^*{\hat a}_-$ and 
$\hbar{\hat b}_+^*{\hat b}_+$: 
\bsub\label{5-8}
\beqn
& &\bbra{c_-}\hbar{\hat a}_+^*{\hat a}_+\kket{c_-}=(|W_-|^2+\hbar)|V_+|^2=K
\ , \label{5-8a}\\
& &\bbra{c_-}\hbar{\hat b}_-^*{\hat b}_-\kket{c_-}
=|W_-|^2+(|W_-|^2+\hbar)|V_+|^2=2(T-\hbar/2)+K \ , 
\label{5-8b}\\
& &\bbra{c_-}\hbar{\hat a}_-^*{\hat a}_-\kket{c_-}
=|A_-|^2=2M-K \ , 
\label{5-8c}\\
& &\bbra{c_-}\hbar{\hat b}_+^*{\hat b}_+\kket{c_-}
=|B_+|^2=2L-K \ . 
\label{5-8d}
\eeqn
\esub
Here, $K$, $T$, $M$ and $L$ denote the expectation values of 
${\wtilde K}$, ${\wtilde T}$, 
${\wtilde M}$ and ${\wtilde L}$ defined in the relation (\ref{3-4}): 
\begin{equation}\label{5-9}
\bbra{c_-}{\wtilde K}\kket{c_-}=K \ , \quad
\bbra{c_-}{\wtilde T}\kket{c_-}=T\ , \quad
\bbra{c_-}{\wtilde M}\kket{c_-}=M\ , \quad
\bbra{c_-}{\wtilde L}\kket{c_-}=L\ . 
\end{equation}
From the relation (\ref{5-8}), we have 
\beqn\label{5-10}
& &|W_-|=\sqrt{2(T-\hbar/2)} \ , 
\qquad |V_+|=\sqrt{{K}/{2T}} \ , \nonumber\\
& &|A_-|=\sqrt{2M-K}\ , \qquad |B_+|=\sqrt{2L-K} \ . 
\eeqn
Since $W_-$, $V_+$, $A_-$ and $B_+$ are complex, we must determine the 
phase angles. 
Then, we regard $K$, $T$, $M$ and $L$ as the action variables and we denote 
their canonically conjugate variables (the angle variables) as $\psi$, 
$\phi_T$, $\phi_M$ and $\phi_L$, respectively. 
This statement is supported by the following condition: 
\beqn\label{5-11}
& &\bbra{c_-}i\hbar\partial_{\psi}\kket{c_-}=K\ , \qquad 
\bbra{c_-}i\hbar\partial_{K}\kket{c_-}=0 \ , \nonumber\\
& &\bbra{c_-}i\hbar\partial_{\phi_T}\kket{c_-}=T-\hbar/2\ , \qquad 
\bbra{c_-}i\hbar\partial_{T}\kket{c_-}=0 \ , \nonumber\\
& &\bbra{c_-}i\hbar\partial_{\phi_M}\kket{c_-}=M\ , \qquad 
\bbra{c_-}i\hbar\partial_{M}\kket{c_-}=0 \ , \nonumber\\
& &\bbra{c_-}i\hbar\partial_{\phi_L}\kket{c_-}=L\ , \qquad 
\bbra{c_-}i\hbar\partial_{L}\kket{c_-}=0 \ . 
\eeqn
The condition (\ref{5-11}) 
was introduced by Marumori, Maskawa, Sakata and Kuriyama for 
choosing collective degrees of freedom in many-fermion system,\cite{18} 
and later, 
the present authors also used extensively in various many-body 
problems.\cite{8,19} 
For the state (\ref{5-3}), we obtain the following formula for any 
variable $z$: 
\beqn\label{5-12}
\bbra{c_-}i\hbar\partial_{z}\kket{c_-}&=&
(i/2)\biggl[(|W_-|^2+\hbar)\left(V_+^*\frac{\partial V_+}{\partial z} 
-V_+\frac{\partial V_+^*}{\partial z}\right)
+\left(W_-^*\frac{\partial W_-}{\partial z} 
-W_-\frac{\partial W_-^*}{\partial z}\right) \nonumber\\
& &\qquad\quad
+\left(A_-^*\frac{\partial A_-}{\partial z} 
-A_-\frac{\partial A_-^*}{\partial z}\right)
+\left(B_+^*\frac{\partial B_+}{\partial z} 
-B_+\frac{\partial B_+^*}{\partial z}\right)\biggl] \ . 
\eeqn
With the aid of the formula (\ref{5-12}), the condition (\ref{5-11}) 
with the relation (\ref{5-10}) leads to the form 
\beqn\label{5-13}
& &W_-=\sqrt{2(T-\hbar/2)}e^{-i\phi_T/2} \ , \qquad
V_+=\sqrt{K/{2T}}e^{-i\psi-i\phi_M/2-i\phi_L/2} \ , \nonumber\\
& &A_-=\sqrt{2M-K}e^{-i\phi_M/2} \ , \qquad
B_+=\sqrt{2L-K}e^{-i\phi_L/2}\ . 
\eeqn

The expectation value of 
$\hbar^2{\hat a}_+^*{\hat b}_-^*{\hat b}_+{\hat a}_-$, 
which characterizes the Hamiltonian (\ref{3-6}) is calculated in the form 
\beqn\label{5-14}
\bbra{c_-}\hbar^2{\hat a}_+^*{\hat b}_-^*{\hat b}_+{\hat a}_-\kket{c_-}
&=&(|W_-|^2+\hbar)U_+V_+^*A_-B_+ \nonumber\\
&=&\sqrt{K(2T+K)(2M-K)(2L-K)}e^{i\psi} \ .
\eeqn
With the use of the relations (\ref{5-8a}), (\ref{5-9}) and 
(\ref{5-14}), we can calculate $\bbra{c_-}{\wtilde H}\kket{c_-}$, 
the result of 
which is reduced to the form (\ref{4-15}). 
Thus, we obtain the classical counterpart of ${\wtilde H}$ in terms of 
calculating the expectation value of ${\wtilde H}$ for the state $\kket{c_-}$. 

In the case (\ref{3-9b}), the state (\ref{3-7}) can be rewritten in the form 
\bsub\label{5-15}
\beqn
\kket{(tml);k}&=&
\sqrt{\frac{(1-2t)!}{(2m-1+2t)!(2l-1+2t)!}}
\sqrt{\frac{((k-1+2t)!)^3}{k!(2m-k)!(2l-k)!}}
\nonumber\\
& &\times ({\hat a}_+^*{\hat b}_-^*{\hat b}_+{\hat a}_-)^{k-1+2t}
\kket{(+)tml} \ , 
\label{5-15a}\\
\kket{(+)tml}&=&
\left(\sqrt{(1-2t)!(2m-1+2t)!(2l-1+2t)!}\right)^{-1}\nonumber\\
& &\times ({\hat a}_+^*)^{1-2t}({\hat a}_-^*)^{2m-1+2t}
({\hat b}_+^*)^{2l-1+2t}\kket{0} \ . 
\label{5-15b}
\eeqn
\esub
The state $\kket{(+)tml}$ obeys 
\begin{equation}\label{5-16}
({\hat a}_-^*{\hat b}_+^*{\hat b}_-{\hat a}_+)\kket{(+)tml}=0 \ . 
\end{equation}
In the same idea as that of the case (\ref{3-9a}), we can 
set up the following wave packet: 
\beqn\label{5-17}
\kket{c_+}&=&
N_c \exp\left(\frac{\hbar}{A_-B_+}\frac{V_-}{U_-}{\hat a}_+^*{\hat b}_-^*
{\hat b}_+{\hat a}_-\right)\nonumber\\
& &\quad\times \exp\left(
\frac{1}{\sqrt{\hbar}}\frac{W_+}{U_+}{\hat a}_+^*\right)
\exp\left(\frac{A_-}{\sqrt{\hbar}}{\hat a}_-^*\right)
\exp\left(\frac{B_+}{\sqrt{\hbar}}{\hat b}_+^*\right)\kket{0} \nonumber\\
&=&
N_c\exp\left(
\frac{V_-}{U_-}{\hat a}_+^*{\hat b}_-^*+\frac{1}{\sqrt{\hbar}}
\frac{W_+}{U_-}{\hat a}_-^*\right)
\cdot\exp\left(
\frac{A_-}{\sqrt{\hbar}}{\hat a}_-^*+\frac{B_+}{\sqrt{\hbar}}{\hat b}_+^*
\right)\kket{0}\ . 
\eeqn
The interpretation of the notations may be not necessary. 
In the same manner as that of the case (\ref{3-9a}), 
which was previously presented, we can formulate 
the mixed-mode coherent state $\kket{c_+}$ and the final result 
is unchanged. 
In this case, the quantity $T$ is negative.

\section{Correspondence between the fermion and the boson representation}

Until the present stage, we describe the two-level pairing model in the 
Schwinger boson representation independently of the model in 
many-fermion system. 
In this section, first, we connect two forms 
with each other. 
For this purpose, let us set up the following correspondence: 
\bsub\label{6-1}
\beqn
& &{\hat {\mib {\cal S}}}(+)^2 \sim {\wtilde {\mib S}}(+)^2 \ , \qquad
{\hat {\mib {\cal S}}}(-)^2 \sim {\wtilde {\mib S}}(-)^2 \ , 
\label{6-1a}\\
& &{\hat {\cal S}}_0(+)\sim {\wtilde S}_0(+) \ , \qquad
{\hat {\cal S}}_0(-)\sim{\wtilde S}_0(-) \ . 
\label{6-1b}
\eeqn
\esub
With the aid of the relations (\ref{2-7}) and (\ref{3-2}), the correspondence 
(\ref{6-1}) leads to 
\bsub\label{6-2}
\beqn
& &\hbar\Omega_+/2 \sim 
(\hbar/2)({\hat a}_+^*{\hat a}_+ + {\hat b}_+^*{\hat b}_+) \ , \nonumber\\
& &\hbar\Omega_-/2 \sim 
(\hbar/2)({\hat a}_-^*{\hat a}_- + {\hat b}_-^*{\hat b}_-) \ , 
\label{6-2a}\\
& &\hbar{\hat {\cal N}}_+ \sim 2\hbar{\hat a}_+^*{\hat a}_+ \ , \nonumber\\
& &\hbar{\hat {\cal N}}_- \sim 2\hbar{\hat a}_-^*{\hat a}_- \ . 
\label{6-2b}
\eeqn
\esub
Combining the relation (\ref{3-4}), we have 
\bsub\label{6-3}
\beqn
& &\hbar\Omega_+/2 \sim {\wtilde L} \ , 
\label{6-3a}\\
& &\hbar{\hat {\cal N}}/4 \sim {\wtilde M} \ , 
\label{6-3b}\\
& &(\hbar/2)(\Omega_- - {\hat {\cal N}}/2+1) \sim {\wtilde T} \ . 
\label{6-3c}
\eeqn
\esub
Here, ${\hat {\cal N}}$ denotes the total fermion number: 
\begin{equation}\label{6-4}
{\hat {\cal N}}={\hat {\cal N}}_+ + {\hat {\cal N}}_- \ .
\end{equation}
Of course, the first relation of the form (\ref{6-2b}) is also 
important: 
\begin{equation}\label{6-3d}
\hbar{\hat {\cal N}}_+/2 \sim {\wtilde K} \ .
\end{equation}
We can learn that ${\wtilde M}$ and ${\wtilde K}$ are closely related to 
the total and the upper level fermion number, respectively. 
In the form of the quantum numbers, we have 
\begin{equation}\label{6-5new}
\Omega_+=2l \ , \qquad \Omega_-=2m+2t-1 \ , \qquad
\nu/2=2m\ . 
\end{equation}
The vacuum $\rket{\Omega_+,\Omega_-}$ corresponds to the following: 
\beqn\label{6-5}
\rket{\Omega_+,\Omega_-}&\sim&\kket{\Omega_+,\Omega_-}=
\left(\sqrt{\Omega_+!\Omega_-!}\right)^{-1}({\hat b}_+^*)^{\Omega_+}
({\hat b}_-^*)^{\Omega_-}\kket{0} \nonumber\\
&=&
\left(\sqrt{(2l)!(2m+2t-1)!}\right)^{-1}({\hat b}_+^*)^{2l}
({\hat b}_-^*)^{2m+2t-1}\kket{0} \ . 
\eeqn

In the fermion space, the state $\rket{\Omega_+,\Omega_-}$ is uniquely 
specified. 
However, in the Schwinger boson representation, 
the state $\kket{\Omega_+,\Omega_-}$ is constructed by successive operation 
of ${\hat b}_+^*$ and ${\hat b}_-^*$. 
Therefore, we can make appropriate superposition of 
$\{ \kket{\Omega_+,\Omega_-};\ \Omega_+,\Omega_-=0,1,2,\cdots \}$, 
for example, 
\begin{equation}\label{6-6}
\kket{v_0}=\exp(\beta_+{\hat b}_+^*)\exp(\beta_-{\hat b}_-^*)\kket{0}\ . 
\end{equation}
Here, $\beta_+$ and $\beta_-$ denote complex parameters. We can see 
that the state $\kket{v_0}$ satisfies the same relation as that for 
$\rket{\Omega_+,\Omega_-}$: ${\hat {\cal S}}_-(+)\rket{\Omega_+,\Omega_-}
={\hat {\cal S}}_-(-)\rket{\Omega_+,\Omega_-}=0$. 
Therefore, in spite of boson number non-conservation, we may expect that 
$\kket{v_0}$ plays the same role as that of $\rket{\Omega_+,\Omega_-}$. 
It should be noted that $\kket{v_0}$ is a kind of the Glauber coherent 
state. 
By operating $\exp(\alpha_+{\wtilde S}_+(+))\exp(\alpha_-{\wtilde S}_+(-))$ 
on the state $\kket{v_0}$, we can set up the state $\kket{c_0}$ as follows: 
\beqn\label{6-7}
\kket{c_0}&=&
N_c\exp(\alpha_+{\wtilde S}_+(+))\exp(\alpha_-{\wtilde S}_+(-))\kket{v_0}
\nonumber\\
&=&N_c\exp[(\hbar\alpha_+\beta_+){\hat a}_+^*
+(\hbar\alpha_-\beta_-){\hat a}_-^*
+\beta_+{\hat b}_+^*+\beta_-{\hat b}_-^*]\kket{0} \ . 
\eeqn
Here, $N_c$ denotes the normalization constant. 
The state $\kket{c_0}$ in the Schwinger boson representation 
is equivalent to the state $\rket{c_0}$ treated in the 
relation (\ref{2-14}). 
It is the Glauber coherent state. 
The detail will be reported in the succeeding paper. 

Next, we investigate the correspondence of the forms (\ref{2-15a}) and 
(\ref{2-15b}). 
First, it should be noted that the states $\rket{(\mp)\Omega_+,\Omega_-}$ 
introduced in the relations (\ref{2-12a}) and (\ref{2-12b}) correspond to 
\bsub\label{6-9}
\beqn
& &\rket{(-)\Omega_+,\Omega_-}
\sim \kket{(-)\Omega_+,\Omega_-}
=\left(\sqrt{(2l)!(2m+2t-1)!}\right)^{-1}
({\hat b}_+^*)^{2l}({\hat b}_-^*)^{2m+2t-1}\kket{0} \ , \nonumber\\
& &
\label{6-9a}\\
& &\rket{(+)\Omega_+,\Omega_-}
\sim \kket{(+)\Omega_+,\Omega_-}
=\left(\sqrt{(2l)!(2m+2t-1)!}\right)^{-1}
({\hat b}_+^*)^{2l}({\hat a}_-^*)^{2m+2t-1}\kket{0} \ . \nonumber\\
& &
\label{6-9b}
\eeqn
\esub
Then, noting the relation $2m=\nu/2$, we have 
\bsub\label{6-10}
\beqn
& &\left({\hat {\cal S}}_+(-)\right)^{\nu/2}\rket{(-)\Omega_+,\Omega_-}
\sim \left({\wtilde S}_+(-)\right)^{2m}\kket{(-)\Omega_+,\Omega_-}
=\kket{(-)tml} \ , 
\label{6-10a}\\
& &\left({\hat {\cal S}}_+(+)\right)^{\nu/2-\Omega_-}
\rket{(+)\Omega_+,\Omega_-}
\sim \left({\wtilde S}_+(+)\right)^{1-2t}\kket{(+)\Omega_+,\Omega_-}
=\kket{(+)tml} \ . \quad\ \ 
\label{6-10b}
\eeqn
\esub
Here, $\kket{(-)tml}$ and $\kket{(+)tml}$ are given in the relations 
(\ref{5-1b}) and (\ref{5-15b}), respectively. 
Further, we have 
\begin{equation}\label{6-8b}
{\hat {\cal S}}_+(+){\hat {\cal S}}_-(-)\sim {\wtilde S}_+(+){\wtilde S}_-(-)
=\hbar^2{\hat a}_+^*{\hat b}_-^*\cdot{\hat b}_+{\hat a}_- \ . 
\end{equation}
Then, we have the correspondence 
\bsub\label{6-12}
\beqn
& &\rket{(\Omega_+\Omega_-\nu);\kappa}\sim \kket{(tml);k} \ , 
\quad (k=\kappa/2) 
\label{6-12a}\\
& &\rket{(\Omega_+\Omega_-\nu);\kappa}\sim \kket{(tml);k} \ . 
\quad (k-1+2t=\kappa/2-(\nu/2-\Omega_-)) 
\label{6-12b}
\eeqn
\esub
The states $\rket{(\Omega_+\Omega_-\nu);\kappa}$ and $\kket{(tml);k}$ in the 
relation (\ref{6-12a}) are defined in the relations 
(\ref{2-12a}) and (\ref{5-1a}), respectively. 
The states $\rket{(\Omega_+\Omega_-\nu);\kappa}$ and $\kket{(tml);k}$ 
in the relation (\ref{6-12b}) are also given in the relations 
(\ref{2-12b}) and (\ref{5-15a}), respectively. 
The above correspondence permits us to set up the states 
which correspond to the states (\ref{2-15a}) and (\ref{2-15b}):
\bsub\label{6-13}
\begin{eqnarray}
& &\kket{c_-}=N_c\exp[\beta{\wtilde S}_+(+){\wtilde S}_-(-)]
\exp(\gamma{\wtilde S}_+(-))\kket{v_-} \ ,
\label{6-13a}\\
& &\kket{c_+}=N_c\exp[\beta{\wtilde S}_+(+){\wtilde S}_-(-)]
\exp(\gamma{\wtilde S}_+(+))\kket{v_+} \ .
\label{6-13b}
\end{eqnarray}
\esub
Here, $\kket{v_-}$ and $\kket{v_+}$ are defined as 
\bsub\label{6-14}
\beqn
& &\kket{v_-}=\exp(\beta_+{\hat b}_+^*)\exp(\beta_-{\hat b}_-^*)\kket{0} \ , 
\qquad 
(\kket{v_-}=\kket{v_0})
\label{6-14a}\\
& &\kket{v_+}=\exp(\beta_+{\hat b}_+^*)\exp(\alpha_-{\hat a}_-^*)\kket{0} \ . 
\label{6-14b}
\eeqn
\esub
Here, $\kket{v_0}$ is defined in the relation (\ref{6-6}). 
The states (\ref{6-13a}) and (\ref{6-13b}) 
are identical with the states given in the 
relation (\ref{5-6}) and (\ref{5-17}), respectively. 
Introduction of the states (\ref{6-14a}) and (\ref{6-14b}) 
and the re-form of 
${\wtilde S}_+(+){\wtilde S}_-(-)$ given in the relation 
(\ref{6-8b}) enable us to present the mixed-mode coherent states (\ref{5-6}) 
and (\ref{5-17}). 
In the original fermion space, it may be impossible to present 
the above treatment.

\section{Discussion and concluding remark}

Finally, as a discussion, we sketch our idea how to describe our system 
based on the Hamiltonian (\ref{4-11}) and its classical counterpart 
(\ref{4-14}) or (\ref{4-15}). 
The detail will be reported in the succeeding paper. 
The simplest idea may be to search the minimum point of the 
energy and the small amplitude oscillation around this point. 
First, we consider the classical case. 
The Hamiltonian (\ref{4-15}) can be symbolically expressed as 
\begin{equation}\label{7-1}
H=F(K)-f(K)+2f(K)(\sin\psi/2)^2\ . 
\end{equation}
Since $f(K)$ is positive definite, the minimum point appears 
at least at the point 
\begin{equation}\label{7-2}
\psi=0\ . 
\end{equation}
Then, the minimum point $K=K_0$ can be found at least at the point 
$K\geq 0$. 
The function $f(K)$ contains $\sqrt{K(2T+K)}$ and $T\geq \hbar/2$. 
This means the following: 
Near the region $K=0$, the function $(F(K)-f(K))$ is a decreasing function, 
and then, afterward, becomes increasing. 
Therefore, we can find the minimum point by the condition that the 
derivative of $(F(K)-f(K))$ for $K$ is equal to zero. 
The case of the Glauber coherent state does not show always such behaviors. 
Under this consideration, $K=K_0$ can be found by the condition 
\begin{equation}\label{7-3}
F'(K_0)-f'(K_0)=0 \ .
\end{equation}
Thus, the Hamiltonian (\ref{4-15}) is approximated as 
\begin{equation}\label{7-4}
H=F(K_0)-f(K_0)+\frac{1}{2}(F''(K_0)-f''(K_0))(K-K_0)^2+
\frac{1}{2}f(K_0)\psi^2 \ .
\end{equation}
Since $(K-K_0)$ and $\psi$ are canonical, the frequency $\omega$ is given by 
\begin{equation}\label{7-4-2}
\omega=\sqrt{(F''(K_0)-f''(K_0))f(K_0)} \ .
\end{equation}
We can see that $(F''(K_0)-f''(K_0))>0$ and $f(K_0)>0$, and then, 
$\omega$ does not vanish. 
In the case of the Glauber coherent state, we know that under certain 
condition
$\omega$ vanishes. 
In \S 1, we mentioned that, in many-body systems such as nuclei, 
we cannot observe sharp phase transition. 
For the confirmation of this point, it may be interesting to 
investigate the behavior of $\omega$. 
We will report it in the succeeding paper.

Next, we investigate the above idea in the framework of quantum theory. 
The Hamiltonian is symbolically expressed as 
\beqn
& &{\hat H}=F({\hat K})-f({\hat K})-(1/2)\left[
\left(\sqrt{\hbar}{\hat c}^*-\sqrt{{\hat K}}\right)\maru{f}({\hat K})
+\maru{f}({\hat K})\left(\sqrt{\hbar}{\hat c}-\sqrt{\hat K}\right)\right] \ , 
\nonumber\\
& &\sqrt{\hat K}\maru{f}({\hat K})=f({\hat K}) \ .
\eeqn
First, we set up the following relation: 
\begin{equation}\label{7-6}
\sqrt{\hbar}{\hat c}=\sqrt{K_0}-i\sqrt{\hbar}{\hat \gamma}\ , \qquad
\sqrt{\hbar}{\hat c}^*=\sqrt{K_0}+i\sqrt{\hbar}{\hat \gamma}^* \ .
\end{equation}
Here, $({\hat \gamma}, {\hat \gamma}^*)$ denotes the fluctuation around the 
equilibrium value $\sqrt{K_0}$ and it is boson operator. 
Our starting assumption is in the condition that 
$(\sqrt{\hbar}{\hat \gamma}, \sqrt{\hbar}{\hat \gamma}^*)$ is of the 
order $\hbar^0$, but small. 
In the framework of the quadratic with respect to the fluctuation, 
we will treat the system. 
The operator $({\hat K}-K_0)$ is expressed in the form 
\begin{equation}\label{7-7}
{\hat K}-K_0=\sqrt{K_0}\cdot i\sqrt{\hbar}({\hat \gamma}^*-{\hat \gamma})
+\hbar{\hat \gamma}^*{\hat \gamma} \ .
\end{equation}
Therefore, under the above assumption, $({\hat K}-K_0)^2$ and 
$\sqrt{\hat K}$ can be 
approximated as 
\begin{eqnarray}
& &({\hat K}-K_0)^2=K_0[i\sqrt{\hbar}({\hat \gamma}^*-{\hat \gamma})]^2 \ , 
\label{7-8}\\
& &\sqrt{\hat K}=\sqrt{K_0}+(1/2)i\sqrt{\hbar}({\hat \gamma}^*
-{\hat \gamma})+(1/8\sqrt{K_0})[(\sqrt{\hbar}({\hat \gamma}^*
+{\hat \gamma}))^2-2\hbar]\ . 
\label{7-9}
\end{eqnarray}
Then, we have 
\begin{equation}\label{7-10}
\sqrt{\hbar}{\hat c}^*-\sqrt{{\hat K}}
=(1/2)i\sqrt{\hbar}({\hat \gamma}^*
+{\hat \gamma})-(1/8\sqrt{K_0})[(\sqrt{\hbar}({\hat \gamma}^*
+{\hat \gamma}))^2-2\hbar]\ .
\end{equation}
With the use of the relation (\ref{7-10}), the following formula is obtained: 
\bsub\label{7-11}
\beqn
(1/2)\left(\sqrt{\hbar}{\hat c}^*-\sqrt{\hat K}\right)\cdot\maru{f}({\hat K})
&=&-(1/16)f(K_0)/K_0\cdot\left[
(\sqrt{\hbar}({\hat \gamma}^*+{\hat \gamma}))^2-2\hbar\right]\nonumber\\
& &+(1/4)\maru{f}(K_0)\cdot i\sqrt{\hbar}({\hat \gamma}^*+{\hat \gamma})
\nonumber\\
& &
-(1/4)\maru{f'}(K_0)\sqrt{K_0}\cdot\hbar({\hat \gamma}^{*2}
-{\hat \gamma}^2+1) \ , 
\label{7-11a}\\
(1/2)\maru{f}({\hat K})\cdot\left(\sqrt{\hbar}{\hat c}-\sqrt{\hat K}\right)
&=&-(1/16)f(K_0)/K_0\cdot\left[
(\sqrt{\hbar}({\hat \gamma}^*+{\hat \gamma}))^2-2\hbar\right]\nonumber\\
& &-(1/4)\maru{f}(K_0)\cdot i\sqrt{\hbar}({\hat \gamma}^*+{\hat \gamma})
\nonumber\\
& &
+(1/4)\maru{f'}(K_0)\sqrt{K_0}\cdot\hbar({\hat \gamma}^{*2}
-{\hat \gamma}^2-1) \ . \qquad
\label{7-11b}
\eeqn
\esub
The relations (\ref{7-8}) and (\ref{7-11}) give us the following 
form for ${\hat H}$: 
\beqn\label{7-12}
{\hat H}&=&
F(K_0)-f(K_0)
+(1/2)(F''(K_0)-f''(K_0))\left[\sqrt{K_0}\cdot i\sqrt{\hbar}
({\hat \gamma}^*-{\hat \gamma})\right]^2 \nonumber\\
& &+(1/2)f(K_0)\left[1/(2\sqrt{K_0})\cdot \sqrt{\hbar}
({\hat \gamma}^*+{\hat \gamma})\right]^2 
+(\hbar/2)(f'(K_0)-f(K_0)/K_0)
\ .\qquad
\eeqn
Comparison of the Hamiltonian (\ref{7-12}) with the classical form (\ref{7-4}) 
is interesting. 
The last term in the $c$-number part comes from the ordering of the 
boson operator $({\hat \gamma}, {\hat \gamma}^*)$. 
Under the present order of the approximation, $(K-K_0)$ and $\psi$ 
are quantized in the form 
\bsub\label{7-13}
\beqn
{\hat K}-K_0&=&\sqrt{K_0}\cdot i\sqrt{\hbar}({\hat \gamma}^*-{\hat \gamma}) 
\ , \label{7-13a}\\
{\hat \psi}&=&
(1/2\sqrt{K_0})\cdot \sqrt{\hbar}({\hat \gamma}^*+{\hat \gamma}) \ .
\label{7-13b}
\eeqn
\esub
The form (\ref{7-13a}) is consistent to the form (\ref{7-7}) and 
commutation relation for $({\hat K}-K_0)$ and ${\hat \psi}$ is given as 
\begin{equation}\label{7-14}
[\ {\hat K}-K_0 \ , \ {\hat \psi}\ ]=-i\hbar \ . 
\end{equation}
Certainly, ${\hat \psi}$ is canonical to $({\hat K}-K_0)$. 
Then, the Hamiltonian (\ref{7-12}) is expressed in the form 
\beqn\label{7-15}
{\hat H}&=&
F(K_0)-f(K_0)
+\hbar\omega{\hat d}^*{\hat d}
+(\hbar/2)(f'(K_0)-f(K_0)/K_0+\omega)
\ . 
\eeqn
Here, $\omega$ is given in the relation (\ref{7-4-2}) and 
$({\hat d} , {\hat d}^*)$ denotes the boson operator which is expressed in 
terms of boson $({\hat \gamma}, {\hat \gamma}^*)$. 
The last term in the Hamiltonian (\ref{7-15}) appears as a result of 
the quantization and it may be interesting to investigate 
its behavior. 

In this paper, we developed a possible method for describing the 
two-level pairing model in the framework of the mean field approximation. 
The idea can be found in the Schwinger boson representation for the 
$su(2)\otimes su(2)$-algebra and the mixed-mode coherent state plays 
a central role. 
In the present framework, we can treat the system in the Glauber coherent 
state, which is equivalent to the BCS theory in the fermion space. 
As a concluding remark, we mention third possibility for the coherent 
state. 
First, we note the correspondence 
\begin{equation}\label{7-16}
\rket{(\Omega_+\Omega_-\nu);\kappa} \sim \kket{(tml);k} \ .
\end{equation}
Both states are given in the relations (\ref{2-12a}) and 
(\ref{3-7}), respectively. 
The state $\kket{(tml);k}$ can be expressed, for example, 
explicitly in the form 
\beqn\label{7-17}
\kket{(tml);k}&=&
({\hat a}_+^*)^k({\hat b}_-^*)^{2t-1+k}({\hat a}_-^*)^{2m-k}
({\hat b}_+^*)^{2l-k}\kket{0} \nonumber\\
&=&({\hat a}_+^*{\hat b}_-^*)^k({\hat a}_-^*{\hat b}_+^*)^{2m-k}
({\hat b}_+^*)^{2(l-m)}({\hat b}_-^*)^{2t-1}\kket{0} \ . 
\eeqn
The form (\ref{7-17}) suggests us the following coherent state: 
\beqn\label{7-18}
& &\kket{c^0}=N_c\exp(\alpha_+{\hat a}_+^*{\hat b}_-^*)
\exp(\alpha_-{\hat a}_-^*{\hat b}_+^*)
\kket{v_0}\ . 
\eeqn
Of course, $N_c$ and $(\alpha_+, \alpha_-, \beta_+, \beta_-)$ denote 
the normalization constant and complex parameters and $\kket{v_0}$ is defined 
in the relation (\ref{6-6}). 
The state (\ref{7-18}) is rewritten as 
\beqn\label{7-19}
\kket{c^0}&=&N_c\exp(\alpha_+{\hat a}_+^*{\hat b}_-^* +\beta_-{\hat b}_-^*)
\exp(\alpha_-{\hat a}_-^*{\hat b}_+^* +\beta_+{\hat b}_+^*)\kket{0} \ .
\eeqn
The state (\ref{7-18}) is closely related to the $su(1,1)$-algebra 
and we call it the squeezed coherent state. 
From the state (\ref{7-19}), we obtain the other three forms of the 
squeezed coherent states under the following replacement: 
\begin{equation}\label{7-21}
\alpha_+ \longleftrightarrow \beta_- \ , \qquad
\alpha_- \longleftrightarrow \beta_+ \ , \qquad
{\hat a}_+^* \longleftrightarrow {\hat b}_-^* \ , \qquad
{\hat a}_-^* \longleftrightarrow {\hat b}_+^* \ . 
\end{equation}
In the succeeding paper, we will report the comparative investigation 
of the Glauber, the mixed-mode and the squeezed coherent state 
including numerical analysis.

Finally, we will give a comment. 
Our disguised form of the two-level pairing model is presented in the 
framework of one kind of boson operator. 
However, this model can be also formulated in terms of the 
$su(1,1)\otimes su(1,1)$-algebra and with the use of two kinds of bosons, 
another disguised form is obtained. 
The details will be reported as a continuation of the present paper. 

\section*{Acknowledgements} 

The present work was started when the authors Y.T. and M.Y. stayed 
at Coimbra in August of 2005 under the invitation from 
Professor J. da Provid\^encia, co-author of this paper. 
They should acknowledge to him for his kind 
invitation. 
Further, the main part was performed when the author J.P. stayed at 
Yukawa Institute for Theoretical Physics, Kyoto University in 
September of 2005 as a visitor. 
He expresses his sincere thanks to Professor T. Kunihiro for his 
kind invitation. 




\end{document}